\documentclass[conference]{IEEEtran}
\IEEEoverridecommandlockouts
\usepackage{cite}
\usepackage{amsmath,amssymb,amsfonts}
\usepackage{graphicx}
\usepackage{textcomp}
\usepackage{xcolor}
\usepackage[utf8]{inputenc}
\usepackage{multirow}
\usepackage{comment}
\usepackage{xspace}
\usepackage{balance}
\usepackage{multirow}
\usepackage{mathtools}
\usepackage{amssymb}
\usepackage{pdfpages}
\usepackage{caption}
\usepackage{subcaption}
\usepackage[export]{adjustbox}
\usepackage{multicol}
\usepackage{amsthm}
\usepackage{float}
\usepackage{url}

\usepackage{algorithm}
\usepackage{algorithmicx}
\usepackage{algpseudocode}
\usepackage{setspace}

\usepackage{gensymb}
\usepackage{booktabs}
\usepackage{balance}

\usepackage{geometry}
\def\BibTeX{{\rm B\kern-.05em{\sc i\kern-.025em b}\kern-.08em
    T\kern-.1667em\lower.7ex\hbox{E}\kern-.125emX}}

\title{High Altitude Platform Stations: the New Network Energy Efficiency Enabler in the 6G Era\thanks{This research was supported  by the Generalitat Valenciana, Spain, through the  CIDEGENT PlaGenT,  Grant CIDEXG/2022/17, Project iTENTE.}}

\author{
  \IEEEauthorblockN{Tailai Song\IEEEauthorrefmark{1},   David Lopez\IEEEauthorrefmark{3},    Michela Meo\IEEEauthorrefmark{1},   Nicola Piovesan\IEEEauthorrefmark{2},   Daniela Renga\IEEEauthorrefmark{1}}
  \IEEEauthorblockA{\IEEEauthorrefmark{1}\small{Politecnico di Torino}, \texttt{first.last@polito.it}}
  \IEEEauthorblockA{\IEEEauthorrefmark{2}\small{Universitat Politècnica de València}, \texttt{dr.david.lopez@ieee.org}}
  \IEEEauthorblockA{\IEEEauthorrefmark{3}\small{Huawei}, \texttt{nicola.piovesan@huawei.com}}
}

\begin{document}

\maketitle

\begin{abstract}
The rapidly evolving communication landscape, with the advent of 6G technology, brings new challenges to the design and operation of wireless networks. One of the key concerns is the energy efficiency of the Radio Access Network (RAN), as the exponential growth in wireless traffic demands increasingly higher energy consumption. In this paper, we assess the potential of integrating a High Altitude Platform Station (HAPS) to improve the energy efficiency of a RAN, and quantify the potential energy conservation through meticulously designed simulations. We propose a quantitative framework based on real traffic patterns to estimate the energy consumption of the HAPS integrated RAN and compare it with the conventional terrestrial RAN. Our simulation results elucidate that HAPS can significantly reduce energy consumption by up to almost 30\% by exploiting the unique advantages of HAPS, such as its self-sustainability, high altitude, and wide coverage. We further analyze the impact of different system parameters on performance, and provide insights for the design and optimization of future 6G networks. Our work sheds light on the potential of HAPS integrated RAN to mitigate the energy challenges in the 6G era, and contributes to the sustainable development of wireless communications.
\end{abstract}

\section{Introduction}
Radio access networks (RANs) are currently experiencing unprecedented growth due to the shift towards fifth generation (5G) technologies, 
which provide high data rates, reduced latency, and increased network capacity. 
However, this new communication era brings new challenges. 
In particular, network densification becomes crucial, 
especially in urban environments where physical, legal, and bureaucratic limitations constrain the installation of new network infrastructure. 
On top of that, concerns about the environmental impact continue to grow. 
The massive deployment of infrastructure, 
with its associated energy consumption, 
is anticipated to contribute significantly to global greenhouse gas emissions, 
thereby accelerating climate change~\cite{shehab20215g,lopezperez2022survey}. 
Consequently, there is a growing interest in developing sustainable and energy-efficient communication technologies for the next generation of wireless networks.

The sixth generation (6G) of wireless networks is expected to revolutionize wireless communication,
by enabling the realization of futuristic applications, 
such as holographic telepresence, autonomous vehicles, and ubiquitous sensing~\cite{pouttu20206g, yang20196g}. 
However, to enable these applications, 
6G networks will require even higher data rates, lower latency, and ultra-reliable and secure communications, 
which will necessitate the deployment of more base stations (BSs) and network elements, 
leading to increased energy consumption~\cite{mahdi20215g}. 
Therefore, it is critical to investigate potential energy-saving opportunities that can be harnessed in the 6G era.

To this end, the integration of aerial BSs into terrestrial networks could play an essential role by providing supplementary capacity and offloading portions of mobile traffic to alleviate the burdens of on-ground BSs. 
Specifically, high altitude platform stations (HAPSs) are promising candidates for self-sustainable network nodes,
as they can be equipped with super macro BS (SMBS)~\cite{alam2021high} capable of providing additional coverage and capacity to the terrestrial network, 
and operate in the stratosphere at an altitude of around 20~km without demanding additional energy consumption from the power grid~\cite{kurt2021vision}. 
Exploiting HAPS-mounted SMBS enables a space-as-a-service paradigm,
to support flexible energy and resource allocation, capacity enhancement, edge computing, and data caching, as well as processing across manifold application domains~\cite{kurt2021vision}. 
Conventionally, HAPSs were regarded mainly as an enhancement for communication service provisioning and were limited to rural areas or catastrophic scenarios, 
where the terrestrial network may not be accessible. 
Meanwhile, only a few studies have investigated the use of HAPS for non-coverage-based applications, 
such as gigabit mobile communications~\cite{shibata2019study}, IoT services~\cite{jia2021toward}, the cooperation with UAVs~\cite{jia2022hierarchical}, and hybrid communication together with satellites~\cite{shah2021adaptive}. Our work aims to fill a gap by identifying the potential benefits of HAPS-integrated RAN,
which supports joint energy and resource allocation strategies,
in network energy efficiency in the 6G era.

The objective of this paper is to assess the energy efficiency of a HAPS-integrated RAN,
by proposing a framework to quantitatively estimate the energy conservation in different scenarios based on real traffic collected in the city of \textit{Milan} and adapted to match the traffic statistics reported by abundant 4G/5G BSs deployed in \textit{China}. 
In particular, our work presents a comprehensive evaluation of the energy-saving potential of a HAPS-integrated RAN through simulations and numerical analysis. 
By leveraging the advantages of HAPS technology, 
we estimate significant energy savings of up to almost 30\%. In general, the novelty of our contribution is twofold: 
\emph{i)} the construction of the system model, 
in which we formalize the capability of the HAPSs to offload part of the terrestrial traffic and that of the network to subsequently deactivate the corresponding unloaded BSs to save energy, 
and \emph{ii)} the parametric analysis, 
in which we inspect the impact of different scenario configurations on the resulting energy conservation.
\section{Problem Statement}
\begin{figure}[t]
    \centering
    \includegraphics[scale=0.33]{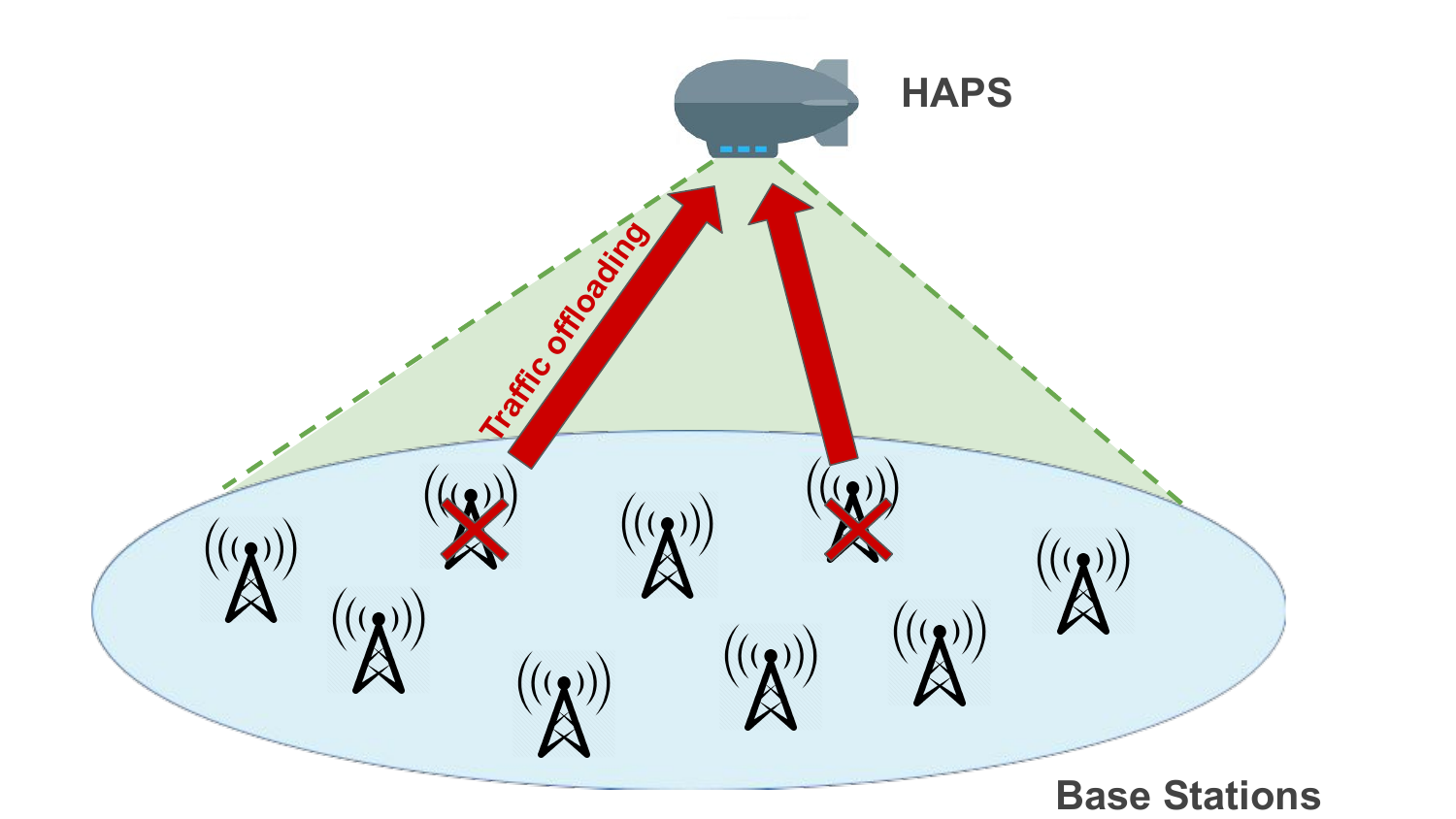}
    \caption{Hybrid aerial-terrestrial Radio Access Network}
    \label{fig:scenario}
\end{figure}

We postulate that the HAPS is deployed and allocated over an urban area of 30~km$^2$ with 960 BSs distributed throughout the region, as shown in Figure~\ref{fig:scenario}. The hourly traffic volumes of each BS during a typical week are considered. Additionally, the HAPS-mounted SMBS is equipped with a $4\times 4$ multiple-input multiple-output (MIMO) radio remote unit (RRU), which can power up to 6 carriers of 20~MHz each. These carriers may be intra-band contiguous. The entire capacity offered by the SMBS is exploited to provide coverage over the entire urban area.

With the objective of reducing the overall energy consumption of terrestrial networks, 
a subset of the terrestrial BSs can be placed in (low-consuming) sleep mode, 
provided that all their traffic can be offloaded to the HAPS. 
To assess the effectiveness of HAPS offloading, we estimate the network energy consumption, satisfying two basic constraints: \emph{i)} the number of active terrestrial BSs cannot be less than a fraction $l_B$ of all terrestrial BSs to preserve quality of service 
(e.g., assuming $l_B=40\%$, 
we can deactivate up to $\lceil 960 \cdot (1-40\%) \rceil = 576$ terrestrial BSs, 
and thus, 
a minimum of $\lceil 960 \cdot 40\% \rceil = 384$ BSs are always active), 
and \emph{ii)} the total offloaded traffic cannot exceed the capacity of the HAPS. 
Denoting by $C_{HAPS}$  the overall HAPS capacity, 
and by $r_{i,h}$ the rate of terrestrial BS $i$  during time step $h$, 
the problem can be formulated as follows:
\begin{center}
\vspace{-2em}
\begin{equation}
\begin{split}
\text{Min} \quad & E_{total}=\sum_{h=1}^{T} \sum_{i=1}^{N} f_{i}(r_{i,h})\times x_{i,h}  \\
\text{s.t.} \quad & \sum_{i=1}^{N} x_{i,h} \geq \lceil N \cdot l_B \rceil, \quad \forall h=1,\ldots,T, \\
& \sum_{i=1}^{N} r_{i,h} \cdot (1-x_{i,h}) \leq C_{HAPS}, \quad \forall h=1,\ldots,T, \\
& x_{i,h} \in {0,1}, \quad \forall i=1,\ldots,N, \quad \forall h=1,\ldots,T, 
\end{split}
\label{eq1}
\end{equation}
\end{center}
where $T$ is the number of time steps during the observation period, 
$N$ is the number of terrestrial BSs, 
$E_{total}$ is the total energy consumed by terrestrial BSs, 
$f_{i}(\cdot)$ is the function that maps the rate of BS $i$ at time step $h$ to its energy consumption, 
and $x_{i,h}$ is a binary decision variable that takes value 1 if BS $i$ is active at time step $h$, 
and 0 if BS $i$ is deactivated with its traffic offloaded to the HAPS, 
and thus, 
$(1-x_{i,h})$ indicates an offloaded BS. 
We aim to minimize $E_{total}$ by selecting the most energy-consuming group of terrestrial BSs 
(i.e., the most appropriate series of $x_{i,h}$), 
offloading the corresponding traffic to the HAPS, 
and putting them into sleep mode.

\section{Methodology}

In this section, 
we model two crucial terms in eq. \eqref{eq1}, namely the function that calculates the energy consumption of terrestrial BSs and the capacity of the HAPS, 
and we propose a simple yet effective traffic offloading algorithm. 
Subsequently, we implement the algorithm in a systematically devised simulation process to estimate the energy saving.

\subsection{System Modelling}
\label{sec:system_modeling}

We first describe the traffic profiles that are adopted in our study.
Then, we detail how the energy consumption of the terrestrial BSs and the HAPS capacity are modeled.

\paragraph{Traffic profiles}
Our simulation is based on real mobile traffic data collected by an Italian telecom operator in the city of Milan in 2015. 
1419 traffic traces, 
corresponding to as many BSs, 
are available, 
each reporting the values of traffic volume on an hourly basis for a period of two months. 
To perform our study in a more realistic scenario, 
hence taking into account the growth of traffic demand observed in the past years, 
the available traffic traces are scaled up, 
matching various aggregated metrics derived from more recent traffic traces that were collected from 960 4G/5G BSs in an urban area in China in 2020. For each of the 960 BSs, 
the peak and the 5$^{th}$ percentile values of the hourly traffic volume, 
computed over a period of one month, 
are provided, 
along with the BS bandwidth capacity and the maximum cell load. 

To obtain updated traffic profiles for each BS in the considered scenario, 
the original traffic profiles from the Italian mobile operator are scaled up according to the following procedure.
First, 
the $N$=1419 original traffic traces are averaged across time to construct a typical week,
thus obtaining $N$ average weekly traces. 
Second, 
for each of the $M$=960 
recent traffic profiles collected from just as many BSs, 
we iterate the following operations. 
A set of $N$ new traces are derived, 
scaling up the original $N$ Italian traces so that each newly derived trace features the same peak and 5$^{th}$~percentile traffic values as the considered $m^{th}$ BS, 
as well as its maximum cell load. 
Furthermore, among the $N$ scaled traffic traces, 
we select the one whose mean value is the closest to the average hourly traffic volume of the considered $m^{th}$ trace. 
At the end of the entire procedure, 
we obtain $M$ new scaled weekly traffic profiles. 
The shape of these newly derived traces remains similar to a subset of the original traffic traces from the Italian mobile operator, 
whereas their peak, 5$^{th}$ percentile, and mean values are scaled to match the corresponding metrics derived from the most recent traffic profiles.
This makes the traffic profiles considered for our investigation up to date, 
and thus more realistic.

\paragraph{Energy consumption model}
We consider the energy consumption model for a 5G BS presented in \cite{piovesan2022machine}.  
This model is utilized as the function $f_{i}(\cdot)$ in eq. \eqref{eq1} to calculate the energy consumption, denoted as $E_{BS}$: 
\begin{equation}
    E_{BS} = E_0 + E_{BB} + E_{Tran} + E_{PA} + E_{out}, \label{eq2}
\end{equation}
where $E_0$ is the baseline energy consumption in sleep mode, 
$E_{BB}$ is the baseband processing energy consumption, 
$E_{Tran}$ is the total energy consumed by the RF chains, 
$E_{PA}$ is the power amplifier (PA) static energy consumption, 
and $E_{out}$ is the energy needed for data transmission. 
The first four parameters are known, 
while the last one is proportional to the transmitted traffic volume: 
\begin{equation}
E_{out}=\frac{1}{\eta} \cdot P_{tx} \cdot \Delta T \cdot \frac{R_{BS}}{C_{BS}}, \label{eq3}
\end{equation}
in which $\eta$ is the efficiency of the PA, 
$P_{tx}$ is the maximum transmit power, 
$\Delta T$ is the duration of the time step, 
and $R_{BS}$ and $C_{BS}$ are the rate and capacity of the corresponding BS,
respectively.
Note that $R_{BS}$ is equivalent to $r_{i,h}$ in eq. \eqref{eq1}. To drive this energy consumption model in a realistic manner, 
in this paper, 
we use the normalized values provided in \cite{piovesan2022machine}.

\paragraph{HAPS capacity model}
\label{haps_c}
The HAPS capacity 
available to offload traffic from terrestrial BSs can be computed based on the Shannon-Hartley theorem:
\begin{equation}
C = B \log_2(1 + \gamma), \label{eq4}
\end{equation}
where $B$ is the bandwidth and $\gamma$ is the Signal-to-Noise Ratio (SNR). 
The SNR can be expressed as:
\begin{equation}
\gamma = \frac{P_{rx}}{N_p}, \label{eq5}
\end{equation}
where $P_{rx}$ denotes the received power and $N_p$ is the noise power. 
Note the HAPSs and the terrestrial BSs operate in a different frequency band,
and thus do not interfere. 

The received power $P_{rx}$ between the HAPS and the user equipment (UE) is calculated as follows \cite{3gpp.38.811, 3gppSimulation}:
\begin{equation}
P_{rx} = P_{tx} + G_{tx} + G_{rx} - PL, \label{eq6}
\end{equation}
where $P_{tx}$ is the HAPS transmit power, 
$G_{tx}$ and $G_{rx}$ are the transmitter and receiver antenna gains, respectively, 
and $PL$ represents the total path loss.

Moreover, 
and for simplicity,
let us assume that the UE benefits from the maximum transmitter antenna gain that the antenna array can provide, 
and thus $G_{tx}$ is defined as follows \cite{5gamericas}:
\begin{equation}
G_{tx} = G_{element} + 10 \log(n\cdot m), \label{eq7}
\end{equation}
where $G_{element}$ is antenna gain of a single element, 
and $n$ and $m$ correspond to the number of rows and columns of antenna elements in the antenna array, respectively.

Additionally, 
the total path loss $PL$ depends on various components, 
according to the following formula \cite{3gpp.38.811}:
\begin{equation}
PL = PL_b + PL_e, \label{eq8}
\end{equation}
where $PL_b$ is the basic path loss and $PL_e$ is the building entry loss. 
Specifically, 
the basic path loss is modeled as:
\begin{equation}
PL_b = FSPL(d,f_c) + SF + CL(\alpha,f_c),\label{eq9}
\end{equation}
where $FSPL(d,f_c)$ represents the free space path loss for a separation distance $d$ in $km$ and frequency $f_c$ in GHz, 
that is given by:
\begin{equation}
FSPL(d,f_c) = 92.45 + 20\log(f_c) + 20\log(d),\label{eq10}
\end{equation} 
$SF$ is the shadow fading loss, 
a random variable characterized by a normal distribution, 
i.e., $SF\sim N(0, \sigma_{SF}^2)$, 
and $CL(\alpha,f_c)$ is the clutter loss, 
with $\alpha$ denoting to the elevation angle. 
Both $\sigma_{SF}^2$ and $CL$ are variables that depend on elevation angles, line-of-sight/non-line-of-sight (LOS/NLOS) conditions\footnote{
The LOS probability is also a function of the elevation angle in different urban scenarios (Table~6.6.1-1 in \cite{3gpp.38.811}).}, 
and frequency ($f_c$, S-band or Ka-band)\footnote{Table~6.6.2-1 in \cite{3gpp.38.811}.}.

The building entry loss $PL_e$ varies depending on the building type, the location within the building and movement in the building. 
The distribution of $PL_e$ is given by a combination of two lognormal distribution
, and given a probability value $P$ ---the maximum loss not exceeded---, 
and, from Equation~(6.6-5) in \cite{3gpp.38.811}, given a probability value $P$ we can compute the maximum loss not exceeded, 
which we denote by $L_{BEL}(P)$.
The model estimates the building entry loss based on two distinct categories of buildings: {\em thermally efficient} and {\em traditional}, 
with the former typically yielding a generally higher entry loss than the latter.

\subsection{Offloading Algorithm}
To solve the minimization problem in eq.~\eqref{eq1},
and since the energy consumption is a function of the cell load, 
we devise an algorithm that attempts to minimize the hourly energy consumption of the terrestrial network through HAPS offloading by prioritizing the offloading from the terrestrial BSs handling the least amount of traffic volume, 
as detailed in Algorithm~\ref{alg:offloading}. 
In this way, 
we can save the non-load dependent energy consumption of more terrestrial BSs. 
Note that $E_{total}$ denotes the overall energy consumption of the terrestrial network during the entire observation period, 
whereas, 
within each time step $h$, 
$n_{offload}$ indicates the current number of offloaded (hence deactivated) terrestrial BSs, 
$R_{offload}$ refers to the overall rate offloaded from the terrestrial BSs to the HAPS, 
and $E_{BS,h}$ is the overall energy consumption of all terrestrial BSs in such time step $h$.
\begin{algorithm}
\caption{The Least Traffic Offloading Algorithm}
\label{alg:offloading}
\begin{algorithmic}[1]
\State \textbf{Input} $R_{BS}$ for each BS, $C_{HAPS}$
\State $E_{total} \gets 0$
\For{$h = 1$ to $T$} 
    \State $n_{offload} \gets 0$
    \State $V_{offload} \gets 0$
    \State $E_{BS,h} \gets 0$
    \State Sort BSs by its rate in ascending order
    \For{each BS in sorted order}
        \If{$n_{offload} \leq \lceil (1-l_B)\cdot N \rceil$ and $R_{offload} \leq C_{HAPS}$}
            \State Offload traffic from BS
            \State $n_{offload} \gets n_{offload} + 1$
            \State $R_{offload} \gets R_{offload} + R_{BS}$
            \State $E_{BS,h} \gets E_{BS,h} + E_0$
        \Else 
            \State Handle traffic with BS
            \State $E_{BS,h} \gets E_{BS,h} +f(R_{BS})$
        \EndIf
    \EndFor
    \State $E_{total} \gets E_{total} + E_{BS,h}$
\EndFor
\State \textbf{Output} $E_{total}$
\end{algorithmic}
\end{algorithm}

\subsection{Simulation Strategy}\label{simu_stra}
At this stage, 
we obtain the necessary information to derive optimized energy savings. 
However, to ensure that our simulation accurately reflects reality, 
we need to consider three parameters: 
\emph{i)} 
\textbf{Elevation angle}: 
In our case, 
the HAPS operates at a height of 20~km with a coverage area of 30~km$^2$. 
Therefore, the elevation angle can be assumed constant within the covered area. 
However, some deviation from the optimal placement directly above the coverage area may be inevitable in reality. 
Hence, we consider an elevation angle that may range between 60° and 90° to account for this uncertainty; 
\emph{ii)} 
\textbf{Building type}: 
As discussed in Section~\ref{sec:system_modeling}, 
different building types contribute to varying levels of building entry loss. 
To generalize our investigation for various scenarios, 
we consider a variable range for the portion of traditional buildings in the urban area under study; 
\emph{iii)}
\textbf{Indoor UE}: 
UEs outside buildings do not suffer from building entry loss, 
and the corresponding calculation of SNR should exclude this factor, 
resulting in a higher value and a larger HAPS rate. 
Hence, we also set a range for the percentage of indoor UEs to account for this variation.

Technically, a larger elevation angle, a higher proportion of traditional buildings, and fewer indoor UEs lead to a larger available HAPS rate, 
hence higher energy conservation. 
However, in order to assess the impact of these parameters, 
a systematic approach is needed to profile the power conservation patterns. 
Therefore, we propose a Monte Carlo simulation process that involves parametric analyses with a certain degree of randomness. 
The randomness includes: 
\emph{i)} the probability of LOS, \emph{ii)} the probability that a UE is indoors, \emph{iii)} the probability that the UE is in a traditional building when the UE is indoors, and \emph{iv)} the probability that the building entry loss will not exceed the corresponding value. 
To ensure statistical significance, 
we perform 1,000 simulation runs, 
each featuring a randomly generated value of elevation angle,  
probability of LOS (that depends on the elevation angle), probability of a UE being indoor, and for indoor UEs, probability of being in a traditional building. 
Furthermore, each simulation deploys 3000 UEs per km$^2$ \cite{earth_d}, 
each characterized by the respective random realization of the various features (LOS/NLOS, indoor/outdoor, traditional/thermally efficient buildings, probability of building entry loss not exceeded), such that the overall distribution of each feature respects the parameter settings of the current trial. The parameters adopted in the simulation are listed in Table~\ref{tab:sim_para}. 

\begin{table}[t]
\centering
\footnotesize
\begin{tabular}{@{}cc@{}}
\toprule
\textbf{Parameter} & \textbf{Value} \\ \midrule
Environment & Dense Urban \\
$l_B$, maximum fraction of offloaded BSs & 0.6 \\
$B$, channel bandwidth {[}MHz{]} & 20 \\
$f_c$, carrier frequency {[}GHz{]} & 2 \\
$d$, HAPS height {[}km{]} & 20 \\
$P_{tx}$, transmit power {[}dBm{]} & 43 \\
$G_{element}$, transmitter gain (single element) {[}dBi{]} & 8 \\
$G_{rx}$, receiver gain {[}dBi{]} & 0 \\
$n$ rows {[}-{]} & 1 \\
$m$ columns {[}-{]} & 4 \\
$T_N$, antenna Noise Temperature {[}K{]} & 290 \\
$N_p$, thermal noise power {[}dBm{]} & -100.96 \\
$\alpha$, elevation angle values & {[}60\degree, 70\degree, 80\degree, 90\degree{]} \\
Portion of traditional building range & (30\%, 70\%) \\
Percentage of indoor user range & (60\%, 90\%) \\ \bottomrule
\end{tabular}
\caption{Simulation parameters}
\vspace{-0.8em}
\label{tab:sim_para}
\end{table}
\section{Experimental Result}

In this section, 
we evaluate the contribution of HAPS offloading to energy savings, 
study the impact on energy conservation of various configuration parameters, 
and provide some insight about HAPS utilization.

\subsection{Energy Saving}

\begin{figure}[t]
    \centering
    \includegraphics[scale=1]{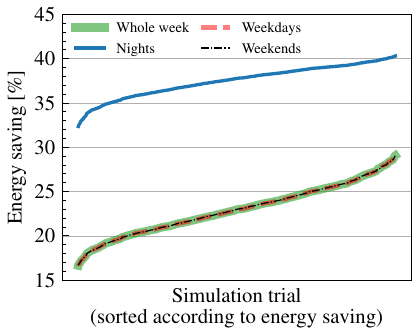}
    \caption{The sorted energy saving for the whole week, nights, weekdays, or weekends.}
    \vspace{-1.6em}
    \label{fig:e_s}
\end{figure}

We define the energy saving as the percentage reduction in energy consumption, 
i.e., the ratio between the energy consumed by all terrestrial BSs under the offloading strategy and the energy consumed by all BSs when no traffic is offloaded. As reported in Figure~\ref{fig:e_s}, 
from the 1,000 performed simulation trials, 
we derive the corresponding values of energy saving, computed over the entire week (green curve), over the weekend (black dotted curve), over the weekdays (red dashed curve), and during the night (0-5 AM, blue curve).

In general, energy savings experience a linear improvement as the trial configuration settings change, regardless of the time period. 
Specifically, significant savings of up to 29\% are achieved throughout the week, 
and even more substantial savings of up to around 41\%  are obtained during the night,
due to lower traffic demand
and the resulting more aggressive offloading. 
Even in the worst-case scenario, 
the HAPS provides energy savings of around 17\% during the week and around 32\% at night. 
Interestingly, the energy conservation pattern for weekdays and weekends coincides with the overall performance, 
resulting in overlapping curves. 
Moreover, despite the fact that savings grow linearly in most cases,  
variations can be observed at the edges, 
especially when it comes to weekly savings. 
This indicates that certain extremely restricted conditions 
(e.g. low elevation angle, huge prevalence of energy efficient buildings, and extremely high portion of indoor users) 
should be prevented to avoid dramatic performance drops. Additionally, it should be noted that from 72 to 159 BSs are never activated during the week depending on the configuration setting. 
This opens the door for deeper sleep modes,
which could be devised to allow even a lower consumption for these BSs,
or a different planning of the terrestrial network might be envisioned to avoid the deployment of unused infrastructure.

\begin{figure}[t]
    \centering
    \includegraphics[scale=0.44]{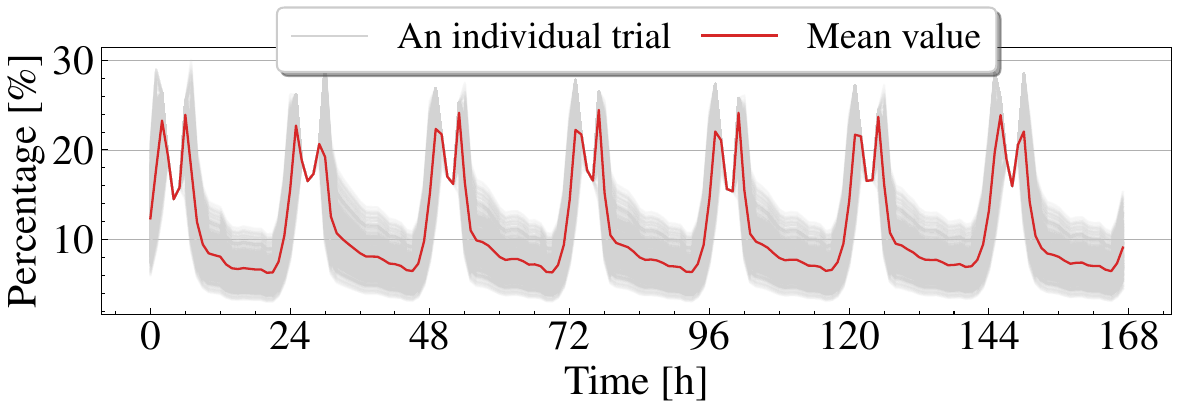}
    \caption{The percentage of offloaded traffic with respect to total traffic.}
    \label{fig:qos}
    \vspace{-1em}
\end{figure}

Another important observation is portrayed in Figure~\ref{fig:qos},
which presents the percentage of offloaded traffic per hour, 
i.e., the ratio between the volume of traffic offloaded to the HAPS and the total traffic demand. 
Specifically, 
it demonstrates that only a handful of traffic is handled by the HAPS, 
especially during the day (3\% to 15\%), 
meaning that most of the traffic is still transmitted by terrestrial BSs. 
Furthermore, despite a limited fraction of traffic, 
ranging from 6.64\% to 16.34\%, 
being offloaded during the whole week on average, 
the energy savings can amount up to 29\%. 
This is due to the important role played by the non-load dependent energy consumption.
Interestingly, 
the amount of traffic handled by the HAPS experiences an abrupt drop during the night. 
This is due to the fact that the threshold for the maximum number of sleeping BSs is quickly reached by extremely under-loaded BSs, 
whereas the capacity of the HAPS is far from saturated, 
given the low total traffic demand. 
Overall, by deactivating a reasonable number of BSs and offloading a small portion of traffic using a limited HAPS capacity, 
we achieve a substantial energy saving, 
which further emphasizes the significance of integrating HAPS into a RAN.
\vspace{-0.2em}

\begin{figure}[t]
    \centering
    \includegraphics[scale=0.5]{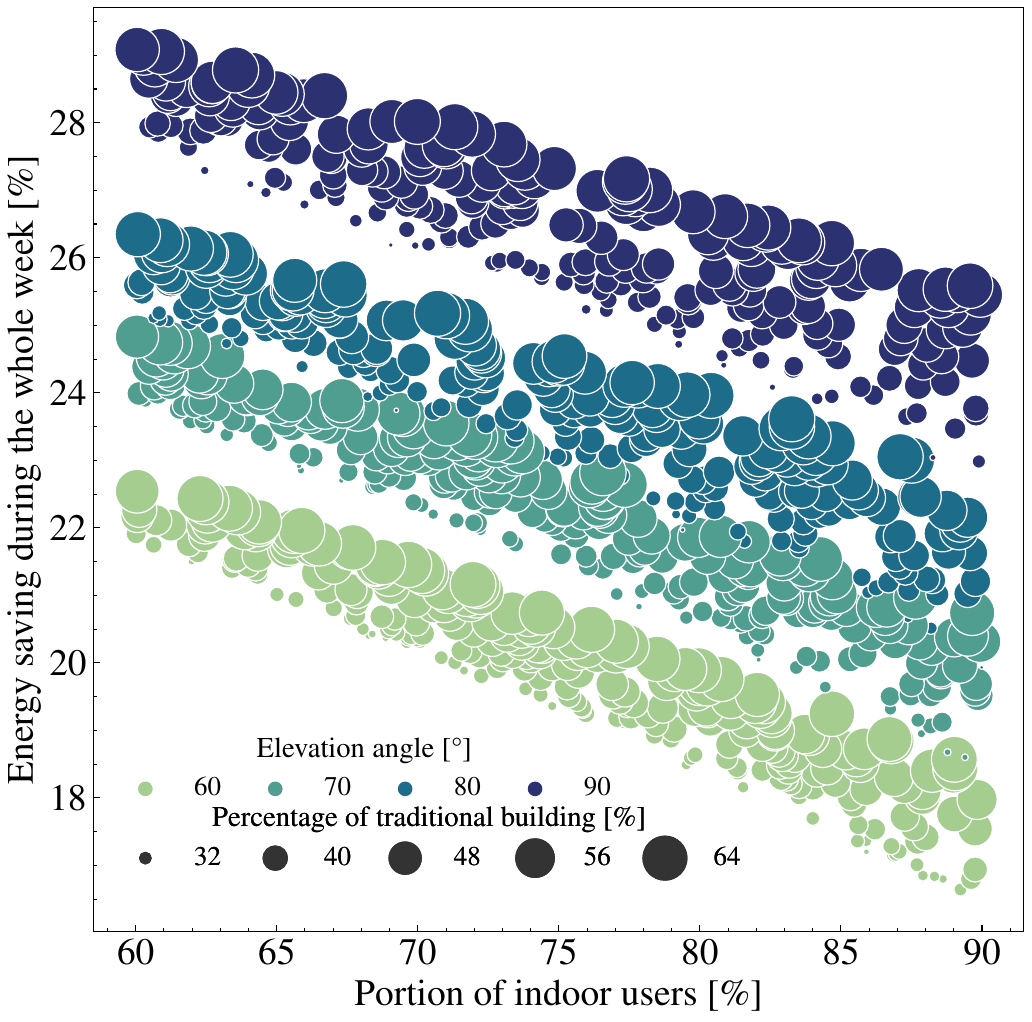}
    \caption{Energy saving at each trial (y-axis) with respect to elevation angle (sample color: the darker the color, the larger the angle), portion of indoor users (x-axis), and percentage of traditional buildings (sample size: the larger the size, the more the traditional buildings).}
    \vspace{-1.6em}
    \label{fig:impact_para}
\end{figure}

\subsection{Impact of Parameters}

As mentioned in Section~\ref{simu_stra}, 
we perform a parametric analysis to evaluate the influence of the elevation angle, the percentage of traditional buildings, and the portion of indoor UEs.
The results are presented in Figure~\ref{fig:impact_para}. 
As expected, 
all parameters show a linear effect on the performance of energy conservation: 
With a larger elevation angle, fewer indoor users, and more traditional buildings, 
we can achieve higher energy saving. 
In particular: 
\emph{i)} The elevation angle has the most distinct impact on energy savings, 
resulting in a stratified pattern in the figure. 
This is due to the fact that the difference in terms of savings between the smallest and largest angles, 
for a given percentage of indoor UEs and traditional buildings, 
is approximately 6 percentage points; 
\emph{ii)} The proportion of indoor UEs shows a less relevant effect on energy savings. 
When the percentage of indoor UEs is around 60\%, 
we can achieve additional savings of approximately 3.6\% with respect to the 90\% of indoor UEs; 
\emph{iii)} The percentage of traditional buildings is the least important factor, 
resulting in a localized effect with respect to elevation angle. 
This leads to a difference of only around 2.5\%. 
Interestingly, as the fraction of indoor UEs increases at a given elevation angle, 
the distribution of energy savings becomes more dispersed with respect to the percentage of traditional buildings, 
indicating an increasingly noteworthy effect when more UEs are inside buildings.

In summary, all parameters show monotonic effects with different degrees of impact. 
To conserve energy, 
the deployment of HAPSs could be advantageous when located right above areas with more people outside and a prevalence of traditional buildings, 
like in a tourist city with historical buildings.
\vspace{-0.1em}

\begin{figure}[t]
    \centering
    \includegraphics[scale=0.5]{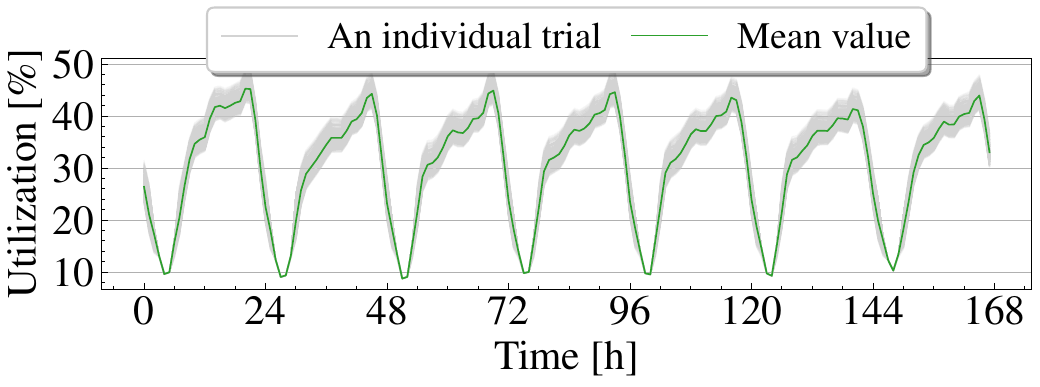}
    \caption{QoS in terms of the utilization of overall available capacity.}
    \label{fig:new_qos}
    \vspace{-1em}
\end{figure}

\subsection{Quality of Service}

When studying power conservation enabled by the HAPS, 
it is important to consider, 
not only the amount of energy saved,
but also the QoS. 
However, measuring QoS at the application level is not feasible in the current scenario where we only have access to hourly traffic volume data. 
As a result, we need to evaluate QoS from a holistic point of view.

From Figure~\ref{fig:qos}, 
it might be evinced that terrestrial BSs, 
which are theoretically more mature and reliable, 
still dominate the communication, 
and due to the fact that only a limited fraction of terrestrial traffic is offloaded to the HAPS regardless of the situations, 
we can reasonably conclude that
the QoS in the HAPSs is equivalent to the one achieved with  the terrestrial RAN only. 
Additionally, the overall utilization of the capacity can serve as an effective indicator. 
Figure~\ref{fig:new_qos} illustrates the utilization of available capacity, 
which is the ratio of the total managed rate to the sum of the capacities of the HAPS and the active terrestrial BSs at each hour during the week in each trial. 
Even during busy hours, 
only between 30\% to 50\% of capacity is exploited, 
and utilization is even lower during the night. 
This represents a higher theoretical available bandwidth per traffic unit that provides a wide margin for additional capacity in case of higher-than-expected traffic demand, 
thus guaranteeing a better QoS.

\section{Conclusion}
In this paper, we propose a simulation-based framework to assess the energy efficiency of the HAPS integrated RAN in the 6G era. Our simulation allows us to profile the energy conservation pattern and quantify the impact of parameters, providing valuable insights into the potential energy savings of different network configurations. The results demonstrate that the proposed solution can enable significant energy savings, with up to 29\% reduction compared to traditional terrestrial networks, and indicate that the integration of HAPS introduces an additional degree of freedom to reduce consumption, leading to sustainable development in the future. Overall, our study highlights the potential of HAPS integrated RAN as a promising solution for energy-efficient wireless networks in the 6G era. We hope that our findings will stimulate further research and development in this area, and contribute to the realization of more sustainable and environmentally friendly wireless networks in the future. Future work is possible for the development of a more robust and efficient offloading strategy that can satisfy complex scenarios with additional constraints.

\balance
\bibliographystyle{ieeetr}
\bibliography{rtc.bib}

\end{document}